\documentclass[aps,pre,superscriptaddress,twocolumn,longbibliography]{revtex4}
\usepackage{graphicx,epstopdf,url}
\usepackage[colorlinks=true,urlcolor=blue,citecolor=blue,linkcolor=blue,urlcolor=blue]{hyperref}
\usepackage[english]{babel}
\usepackage[active]{srcltx}
\usepackage{multirow}
\begin{document}

\title{
Influence of a graphene substrate on the stabilization of molecular systems with hydrogen bonds
}
\author{Alexander V. Savin}
\email{asavin@chph.ras.ru}
\affiliation{
N.N. Semenov Federal Research Center for Chemical Physics of the Russian Academy of Sciences,
4 Kosygin St., Moscow 119991, Russia}
\affiliation{
Plekhanov Russian University of Economics, 36 Stremyanny Lane, Moscow 117997, Russia
}

\begin{abstract}
Numerical simulation of the dynamics of planar two- and three-layer molecular structures formed by $\beta$-sheets of polyglycine peptide chains and systems of parallel Kevlar (para-aramid) molecules placed on a graphene sheet has been performed.
It is shown that in these structures the $\beta$-sheets retain their shape, due to the presence of parallel chains of hydrogen bonds, up to a temperature of $T=800$~K.
An even higher stability is exhibited by the system of parallel Kevlar molecules. Here, the parallel chains of hydrogen bonds between peptide groups of neighboring molecules are preserved even at higher temperatures.
The performed modeling allows us to conclude that the addition of graphene to Kevlar fibers can significantly increase their thermal stability.
\end{abstract}

\maketitle

\section{Introduction}
\label{Introduction}

The structure of many molecular systems is characterized by the presence of hydrogen bond chains.
Such structures include $\alpha$-helices and $\beta$-sheets of proteins, DNA double helices, organic crystals of acetamides (e.g., acetanilide (CH$_3$CONHC$_6$H$_5$)$_x$ \cite{Careri1984,Eilbeck1984}), aramid crystals (Kevlar \cite{Chowdhury2018}), proton transport channels in cell membranes \cite{Nagle1983,Kreuer1996,Kaliman2008,Nagamani2011}, crystals of alcohols, hydrogen halides \cite{Jansen1987,Springborg1988}, as well as many other compounds.

Molecular complexes with hydrogen bonds are characterized by a relatively low stabilization energy (0.12--0.40 eV) per hydrogen bond, which ensures their high dynamic mobility.
Such systems have a complex multicomponent structure; they consist of nonlinearly interacting subsystems (electronic, excitonic, phononic, ionic, protonic).
In these systems, due to the mutual compensation of nonlinear effects and dispersion, stable solitary waves (solitons, kinks, breathers, \dots) can be formed.
With their participation, highly efficient transfer of information, energy, and charges can occur along hydrogen bond chains.

The $\alpha$-helix of a protein is stabilized by three parallel chains of HNCO peptide groups (PGs) with hydrogen bonds:
\begin{equation}
{\cdots}{\rm H{-}N{-}C{=}O}{\cdots}{\rm H{-}N{-}C{=}O}{\cdots}{\rm H{-}N{-}C{=}O}{\cdots},
\label{f1}
\end{equation}
where lines denote valence bonds and dots denote hydrogen bonds.
According to Davydov's theory, energy released during the hydrolysis of an adenosine triphosphate (ATP) molecule can be transported along these chains in the form of autolocalized states of amide-I vibrations of the double valence bond C=O, which are part of the PGs.
Davydov and Kislukha as early as 1973 \cite{Davydov1973} (see also \cite{Davydov1976,Davydov1979,Davydov1981}), based on the regularity of the helical structure of proteins, proposed a model of autolocalization of the amide-I vibration due to its nonlinear interaction with longitudinal deformations of hydrogen bond chains (\ref{f1}).
Using this model, it was shown that nonlinear collective autolocalized excitations, which later became known as Davydov solitons, can propagate without energy loss and shape change in $\alpha$-helical protein molecules \cite{Hyman1981,Scott1982,Scott1985,Scott1992}.
The Davydov model is still actively used -- see works \cite{Georgiev2020,Georgiev2022,Cruzeiro2016,Cruzeiro2020,Cruzeiro2022}.

The presence of hydrogen bond chains (\ref{f1}) also allows the transfer of an external electron in $\alpha$-helical proteins in the form of an electrosoliton (a bound state of a localized electron with a region of helix deformation arising from the interaction of the electron with PG chains) \cite{Davydov1979a,Davydov1984,Davydov1991}.

Almost all molecular complexes with hydrogen bond chains become unstable at high temperatures (at $T>100^\circ$C all hydrogen bond chains are disrupted).
This complicates the use of such molecular systems in nanotechnology.
Placing molecular systems containing hydrogen bond chains inside carbon nanotubes leads to their additional stabilization.
Here, hydrogen bond chains can remain stable even at a temperature of 500~K \cite{Savin2020,Savin2024}.
In the present work, we will show that using complexes of such molecular systems with graphene sheets also allows one to significantly increase the sta\-bi\-lity of hydrogen bond chains.
To do this, two-dimensional systems ($\beta$-sheet of a protein, layers of Kevlar molecules) should be placed on graphene sheets of appropriate size.
In this case, the hydrogen bond chains will retain their structure, and the interaction with the graphene sheet substrate will lead to their additional stabilization, which will allow them to be used for energy and charge transfer over a wider temperature range.
In such composite systems, hydrogen bond chains can remain stable even at $T=600$~K.

\section{$\beta$-sheet of a protein}

To analyze the stability of a $\beta$-sheet, consider a poly\-glycine chain (Gly)$_N$ consisting of $N$ peptide groups.
The chemical formula of such a chain is
\begin{equation}
{\rm C}_\alpha{\rm H}_3{\rm-HNCO{-(}C}_\alpha{\rm H}_2{\rm{-HNCO-)}}_{N-2}{\rm C}_\alpha{\rm H}_2{\rm-H}_2{\rm NCO},
\label{f2}
\end{equation}
where two hydrogen atoms are attached to each $\alpha$-carbon atom (so that there are no free valencies, three hydrogen atoms are attached to the first $\alpha$-carbon atom, and one hydrogen atom is attached to the nitrogen atom of the last peptide group).

When modeling the polyglycine chain (\ref{f2}), it is convenient to consider the $\alpha$-carbon atoms with hydrogen atoms attached to them as one united atom $C_\alpha$.
In this case, each chain link will consist of five atoms C$_\alpha$, C, O, N, H with coordinates $\{ {\bf u}_{n,i}\}_{i=1}^5$, where $n$ is the link number of the polypeptide chain and $i$ is the atom number.
Atomic masses: $M_{n,1}=14m_p$, $M_{n,2}=12m_p$, $M_{n,3}=16m_p$, $M_{n,4}=14m_p$, $M_{n,5}=m_p$
(for end links $M_{1,1}=15m_p$, $M_{N,4}=15m_p$), where $m_p=1.6603\times 10^{-27}$~kg is the proton mass.
To describe the interatomic interaction, we use the general force field for organic molecules AMBER (version 2.1, April 2016) \cite{Amber}.
The Hamiltonian of the polypeptide chain is
\begin{equation}
H_{\rm p}=\sum_{n=1}^N\sum_{i=1}^5\frac12 M_{n,i} (\dot{\bf u}_{n,i}, \dot{\bf u}_{n,i}) + P(\{ {\bf u}_{n,i}\}_{n=1,i=1}^{N,~5}),
\label{f3}
\end{equation}
where the first sum defines the kinetic energy (the three-dimensional vector ${\bf u}_{n,i}=(x_{n,i},y_{n,i},z_{n,i})$ defines the coordinates of the $i$-th atom of the $n$-th chain link), and the last term is the potential energy of the chain found using the AMBER force field.

The Hamiltonian of a graphene sheet is
\begin{equation}
H_{\rm c}=\sum_{n=1}^{N_{\rm g}}\left[\frac12M_{n,0}(\dot{\bf v}_n,\dot{\bf v}_n)+Q_n+W(z_n)\right],
\label{f4}
\end{equation}
where $N_{\rm g}$ is the number of atoms in the sheet, $M_{n,0}$ is the mass of the $n$-th carbon atom, and the vector ${\bf v}_n(t)=(x_n,y_n,z_n)$ defines its coordinates at time $t$.
We assume that hydrogen atoms are attached to the edge atoms of the sheet, and the CH group will be considered as a single united atom with mass $M_{n,0}=13m_p$ (the mass of internal atoms is $M_{n,0}=12m_p$).

The term $Q_n$ defines the interaction energy of the $n$-th atom with neighboring atoms of the sheet (accounting for deformation of valence bonds, valence and torsion angles).
A detailed description of the force field used is given in works \cite{Savin2010,Savin2017}.
The potential
\begin{equation}
W(z)=\epsilon_z[m(z_0/z)^l-l(z_0/z)^m]/(l-m),
\label{f5}
\end{equation}
describes the interaction of sheet atoms with a stationary flat substrate (half-space $z\le 0$) on which the sheet is located.
Here $\epsilon_z$ is the interaction energy of a sheet atom with the substrate, $z_0$ is the equilibrium distance from the atom to the substrate plane, and the exponent $l>m$.
For the surface of an h-BN crystal exponents are $l=10$, $m=3.75$, the energy is $\epsilon_z=0.0903$~eV, and the distance is $z_0=3.46$~\AA~\cite{Savin2021a,Savin2021b}.
To model a free graphene sheet, the last term in the Hamiltonian (\ref{f4}) should be removed.

The interaction of polypeptide chain atoms with graphene sheet atoms is described using Lennard-Jones potentials with parameters from the AMBER force field.

The simplest polypeptide chain (Gly)$_N$, in addition to helical conformations, also allows the existence of a planar structure -- a $\beta$-sheet ($\beta$-strand) consisting of a system of parallel hydrogen bond chains between peptide groups.
Such a planar structure, unstable in three-dimensional space, becomes stable if placed on a flat substrate provided by a graphene sheet.
\begin{figure}[tb]
\begin{center}
\includegraphics[angle=0, width=1.0\linewidth]{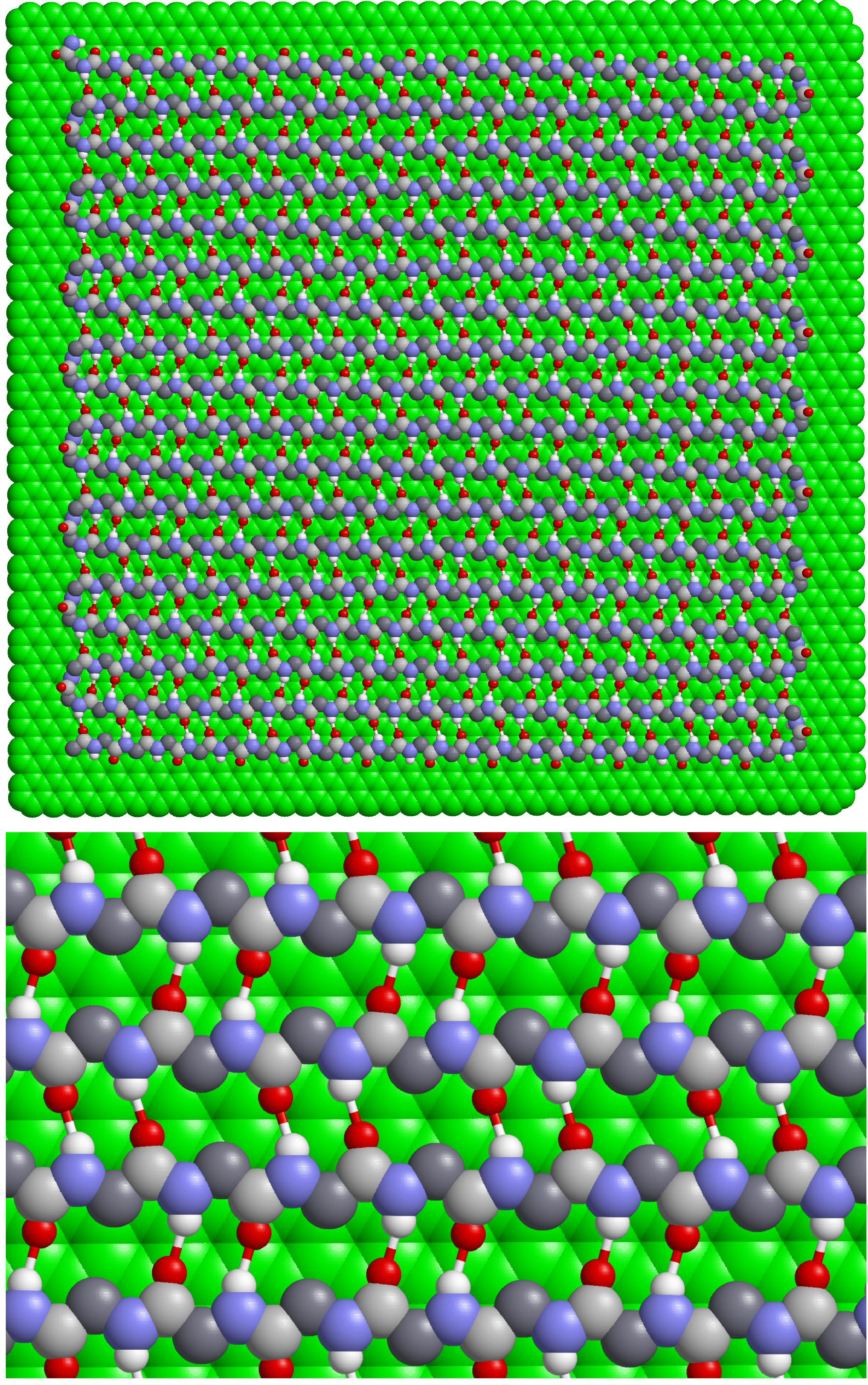}
\end{center}
\caption{\label{fig01}\protect
Stationary state of an antiparallel $\beta$-sheet of the polypeptide chain (Gly)$_{432}$ on a graphene sheet G=C$_{3518}$ of size $9.7\times 9.2$~nm$^2$ (two-layer complex (Gly)$_N$/G$|$ with $N=432$, $N_{\rm g}=3518$, number of hydrogen bonds $N_{\rm hb}=391$).
Graphene sheet atoms are shown in green, peptide group carbon atoms in light gray,
$\alpha$-atoms of the peptide chain in dark gray, nitrogen atoms in blue, oxygen in red, hydrogen in white.
Hydrogen bonds between peptide groups are shown as thin red-white lines.
The bottom shows an enlarged central part of the molecular system.
}
\end{figure}

For definiteness, we take a rectangular graphene sheet G=C$_{N_{\rm g}}$ with a ``zigzag'' structure along the $x$ axis and an ``armchair'' structure along the $y$ axis.
We place on the sheet an antiparallel $\beta$-sheet of the polypeptide chain (Gly)$_N$ in which neighboring hydrogen bond chains are directed in opposite directions -- see Fig. \ref{fig01}.
Antiparallel sheets are more stable than parallel sheets, in which all hydrogen bonds are directed in the same direction.
Such a two-layer structure can be denoted by the formula (Gly)$_N$/G$|$, where the vertical line denotes a solid flat substrate on which the graphene sheet G lies.
We also consider an isolated three-layer structure (Gly)$_N$/G/(Gly)$_N$, in which two identical polypeptide chains are located on different surfaces of the sheet.

To find the stationary state of the two-component system (Gly)$_N$/G$|$, one must solve the problem of minimizing its potential energy
\begin{equation}
E\rightarrow\min: \{ {\bf v}_n\}_{n=1}^{N_{\rm g}},\{ {\bf u}_{n,i}\}_{n=1,i=1}^{N,~5}.
\label{f6}
\end{equation}
We take a rectangular graphene sheet of size $17.56\times 18.58$~nm$^2$ consisting of $N_{\rm g}=12670$ carbon atoms (G=C$_{12670}$), and a polypeptide chain of $N=1584$ PGs.
Problem (\ref{f6}) was solved numerically using the conjugate gradient method \cite{Fletcher1964,Shanno1976}.
The solution of the problem showed that in the ground stationary state, the polypeptide chain lies in a plane parallel to the graphene sheet at a distance $h\approx 3.33$~\AA.
In the most energetically favorable packing, the $\beta$-sheet has the shape of a rectangle of size $15.9\times 16.7$~nm$^2$.
Here, the polypeptide chain is folded in a zigzag, forming $N_1=36$ parallel lines.
This structure is stabilized by $N_2=43$ transverse hydrogen bond chains (the number of hydrogen bonds $N_{\rm hb}=N_2(N_1-1)=1505$).
A structure two times smaller with $N=432$, $N_{\rm g}=3518$ ($N_1=18$, $N_2=23$, $N_{\rm hb}=391$) is shown in Fig. \ref{fig01}.
The average hydrogen bond energy is $E_{\rm hb}=0.304$~eV.
\begin{figure}[tb]
\begin{center}
\includegraphics[angle=0, width=1.0\linewidth]{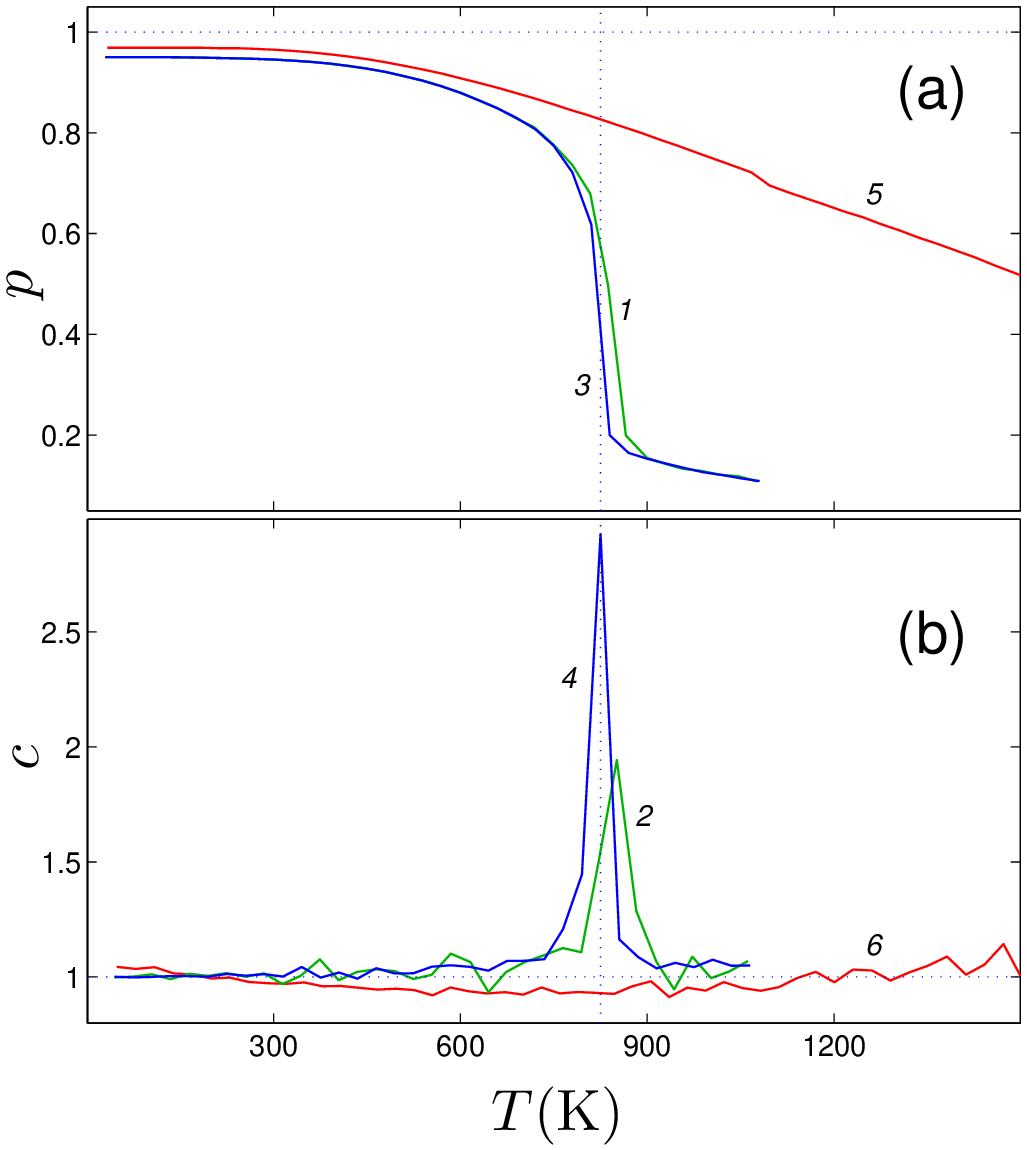}
\end{center}
\caption{\label{fig02}\protect
Dependence (a) of the fraction of peptide groups participating in the formation of hydrogen bonds $p$, and (b) the dimensionless heat capacity $c$ for two- and three-layer complexes (Gly)$_N$/G$|$ (curves 1, 2) and (Gly)$_N$/G/(Gly)$_N$ (curves 3, 4), number of PGs $N=1584$, number of graphene sheet atoms $N_{\rm g}=12670$, and also for the three-layer complex (Aramid$_N$)$_{N_1}$/G/(Aramid$_N$)$_{N_1}$ (curves 5, 6), number of molecules $N_1=32$, number of PGs per molecule $N=24$.
The vertical line shows the temperature $T_0=825$~K at which the $\beta$-sheet is destroyed.
}
\end{figure}

To model the dynamics of two- and three-layer structures (Gly)$_N$/G$|$ and (Gly)$_N$/G/(Gly)$_N$, we first find the ground stationary state of the molecular system by numerically solving the problem of minimizing its potential energy (\ref{f6}).
Then we place the resulting stationary state of the molecular system into a Langevin thermostat and obtain its thermalized state.
To do this, we numerically integrate the Langevin equation system
\begin{equation}
\label{f7}
{\bf M}\ddot{X}=-\frac{\partial H}{\partial X}-\Gamma {\bf M}\dot{X}-\Xi ,
\end{equation}
with the initial condition corresponding to the stationary state of the molecular structure.
Here: $H$ is the total Hamiltonian of the system; $X=(\{ {\bf v}_n\}_{n=1}^{N_g},\{ {\bf u}_{n,i}\}_{n=1,i=1}^{N,~5})$ is the $3N_a$-dimensional vector of atomic coordinates of the molecular structure ($N_a=N_{\rm g}+5N$ is the total number of atoms); ${\bf M}$ is the\linebreak $3N_a$-dimensional vector of atomic masses; $\Gamma=1/t_r$ is the friction coefficient characterizing the intensity of energy exchange with the thermostat (relaxation time $t_r=10$~ps); $\Xi=\{\xi_{n,i}\}_{n=1,i=1}^{N_a,~3}$ is the $3N_a$-dimensional vector of normally distributed random forces normalized by the conditions
\[
\langle\xi_{n,i}(t_1)\xi_{k,j}(t_2)\rangle=2M_n k_B T\Gamma\delta_{nk}\delta_{ij}\delta(t_2-t_1)
\]
($T$ is the thermostat temperature, $k_B$ is the Boltzmann constant).

The system of equations of motion (\ref{f7}) was integrated numerically using the velocity Verlet method \cite{Verlet1967} with an integration step of $\Delta t= 1$~fs.

After a time $t=10t_r$, the molecular system reaches equilibrium with the thermostat at temperature $T$.
Further integration of the system of equations of motion (\ref{f7}) over a time of $t=2$~ns allows us to find the average energy of the system $\bar{E}(T)$ and the average number of hydrogen bonds $\bar{N}_{\rm hb}(T)$ (we assume that two peptide groups form a hydrogen bond if their interaction energy $E>0.18$~eV).
\begin{figure}[tb]
\begin{center}
\includegraphics[angle=0, width=1.0\linewidth]{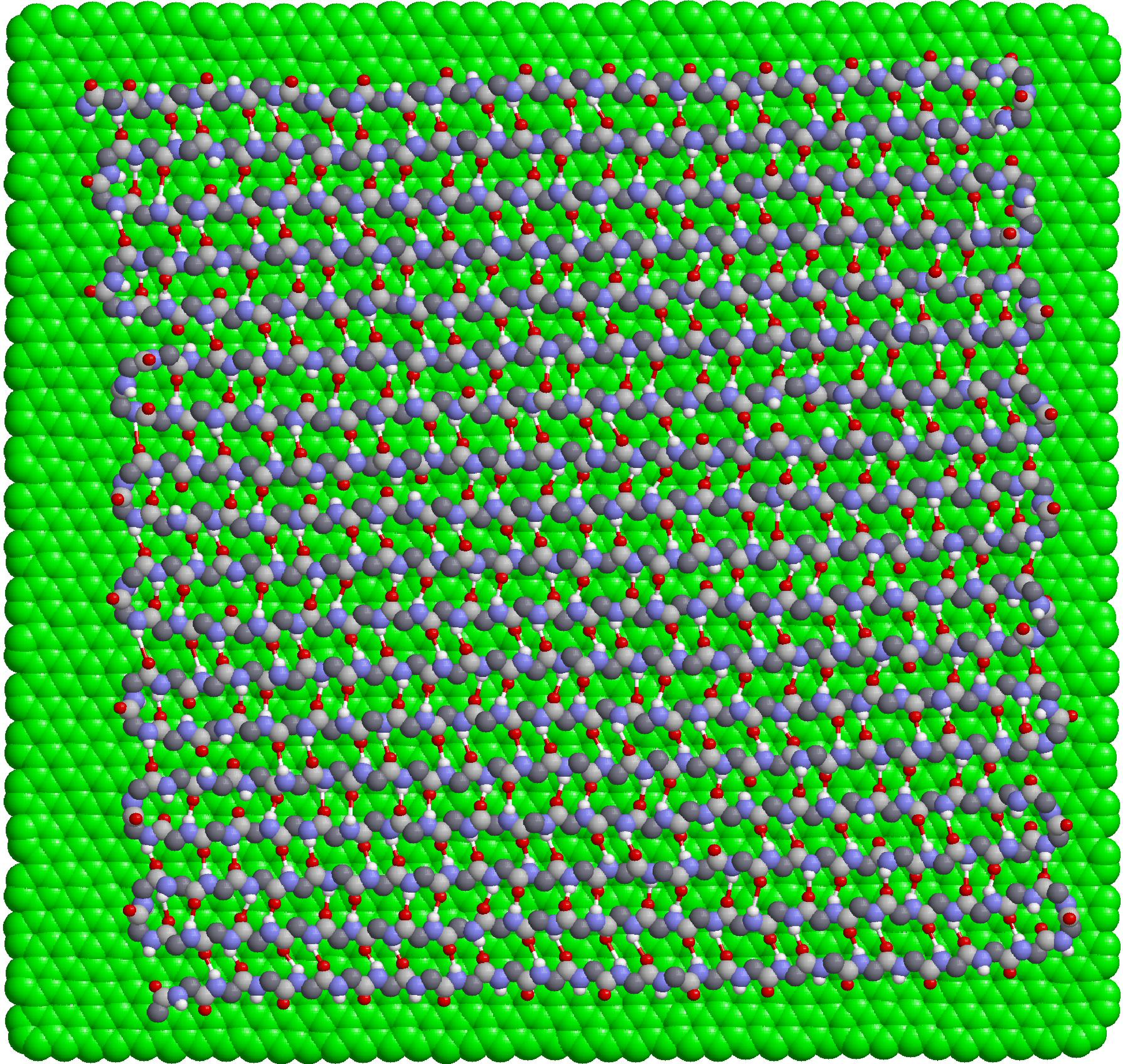}
\end{center}
\caption{\label{fig03}\protect
State of the $\beta$-sheet of the polypeptide chain (Gly)$_{432}$ on a graphene sheet G=C$_{3518}$ (two-layer complex (Gly)$_N$/G$|$ with $N=432$, $N_g=3518$) at temperature $T=600$~K.
$9\%$ of the hydrogen bonds are broken (weakened) ($\bar{N}_{\rm hb}=356.2$).
}
\end{figure}
\begin{figure}[tb]
\begin{center}
\includegraphics[angle=0, width=1.0\linewidth]{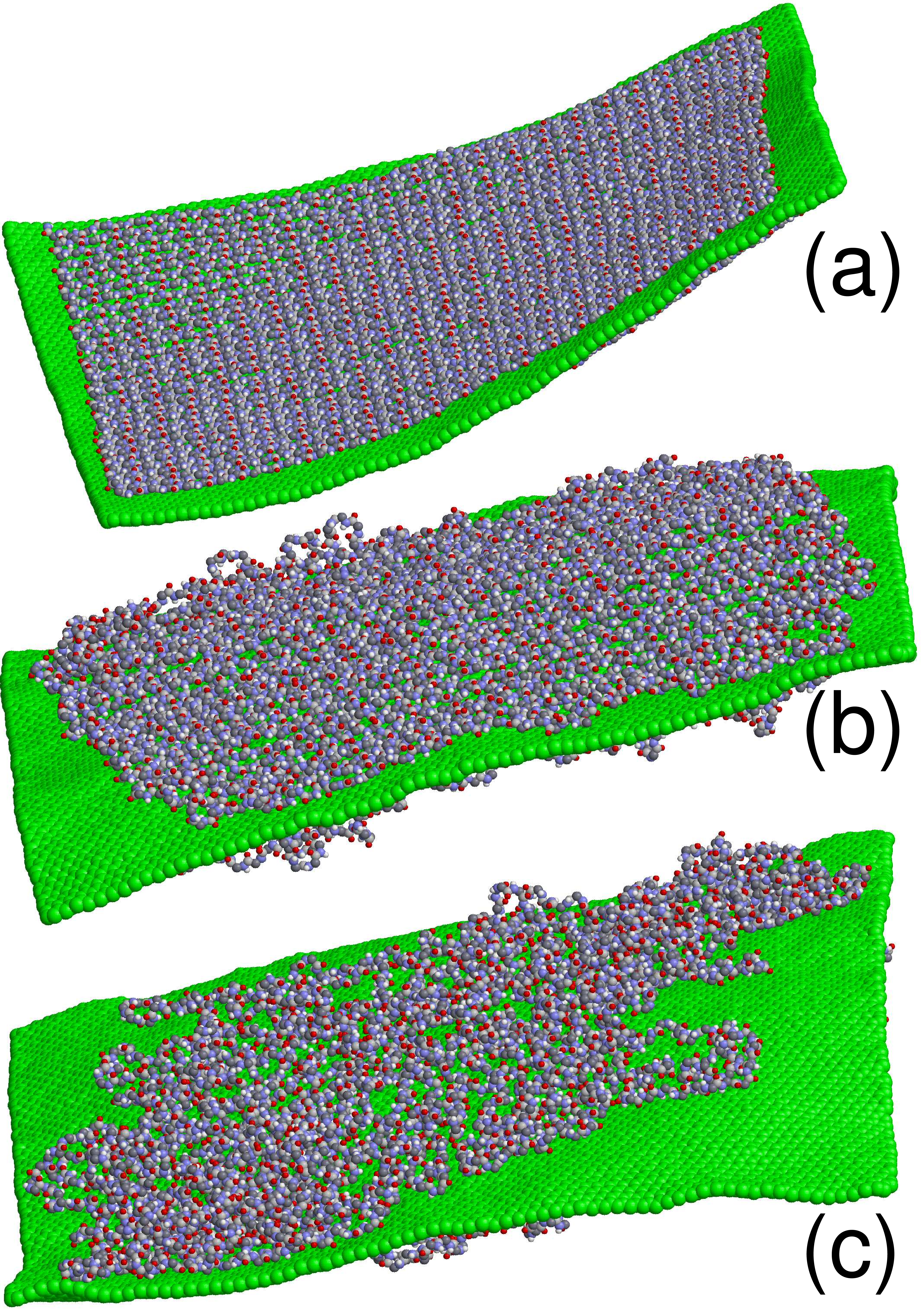}
\end{center}
\caption{\label{fig04}\protect
State of an isolated three-layer complex (Gly)$_N$/G/(Gly)$_N$ ($N=1584$, $N_{\rm g}=12670$) at temperatures (a) $T=300$, (b) 810 and (c) 840~K.
}
\end{figure}

The state of a molecular system can be characterized by its dimensionless heat capacity
\begin{equation}
c=\frac{1}{3N_ak_B}\frac{d\bar{E}(T)}{dT}
\label{f8}
\end{equation}
and the fraction of peptide groups participating in the formation of hydrogen bonds
\begin{equation}
p=\bar{N}_{hb}(T)/N,
\label{f9}
\end{equation}
$p\in [0,1]$, with $p=1$ meaning that each PG participates in the formation of two hydrogen bonds.
The dependence of these quantities on temperature for two- and three-layer complexes (Gly)$_N$/G$|$ and (Gly)$_N$/G/(Gly)$_N$ is shown in Fig. \ref{fig02}.

Numerical integration of the system of equations of motion (\ref{f7}) showed that the $\beta$-sheet of the polyglycine chain (Gly)$_N$ placed on a graphene sheet retains its shape at temperatures $T<T_0=825$~K.
With increasing temperature, only weakening (breaking) of some hydrogen bonds occurs.
For the stationary state of the two-layer structure (at $T=0$), the fraction of PGs participating in the formation of hydrogen bonds is $p=0.950$.
A noticeable decrease in the number of hydrogen bonds occurs only at $T>300$~K -- see Fig.~\ref{fig02}~(a).
\begin{figure}[tb]
\begin{center}
\includegraphics[angle=0, width=1.0\linewidth]{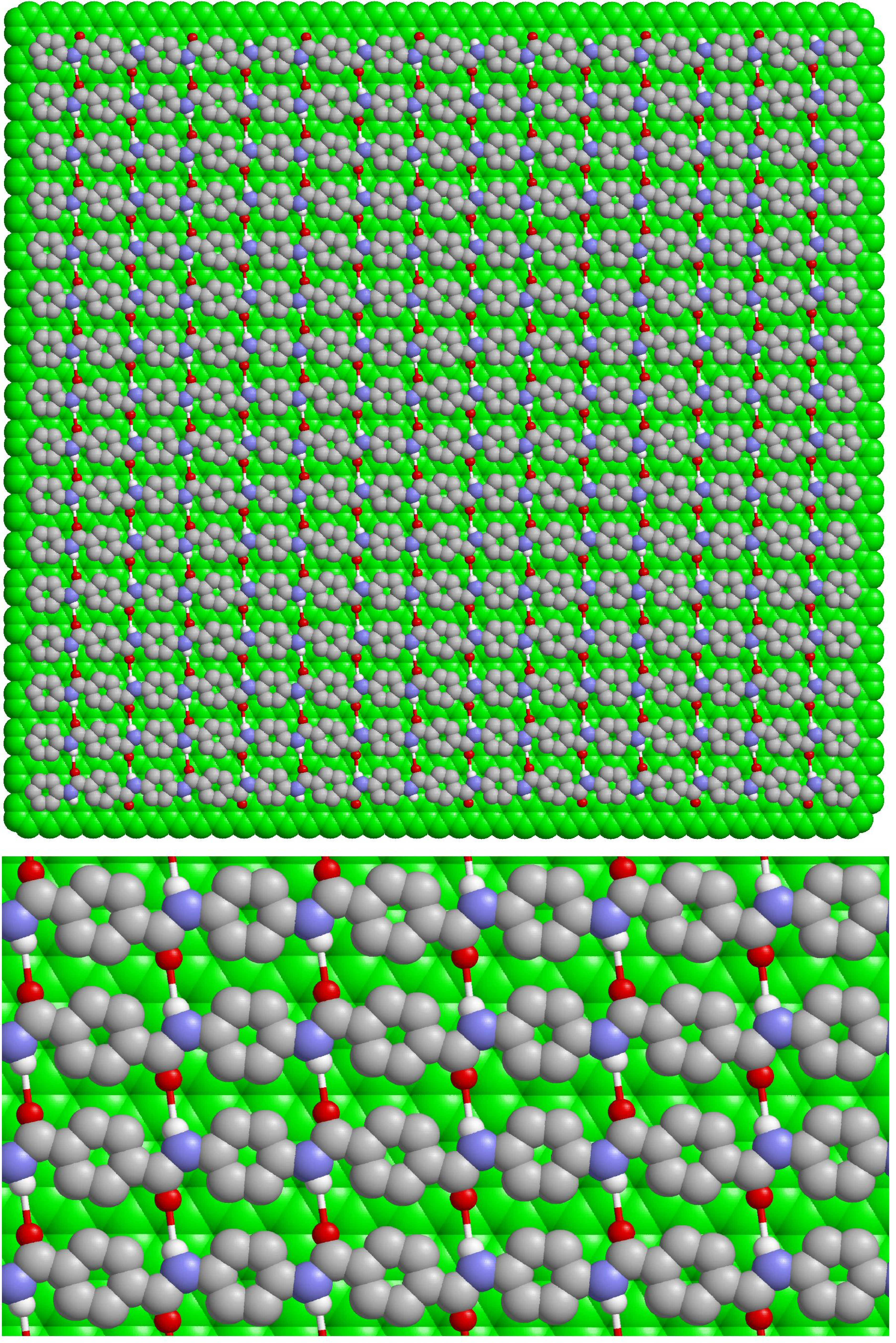}
\end{center}
\caption{\label{fig05}\protect
Stationary state of a system of $N_1=16$ parallel para-aramid molecules H(C$_6$H$_4$CONH)$_{14}$C$_6$H$_5$ lying on a graphene sheet G=C$_{3518}$ of size $9.7\times 9.2$~nm$^2$ (two-layer complex (Aramid$_N$)$_{N_1}$/G$|$ with $N=14$, $N_1=16$, $N_{\rm g}=3518$, number of hydrogen bonds $N_{\rm hb}=210$).
Graphene sheet atoms are shown in green, carbon atoms of para-aramid molecules in gray, nitrogen atoms in blue, oxygen in red, hydrogen in white.
Hydrogen bonds between peptide groups are shown as thin red-white lines.
The bottom shows an enlarged central part of the molecular system.
}
\end{figure}
\begin{figure}[tb]
\begin{center}
\includegraphics[angle=0, width=1.0\linewidth]{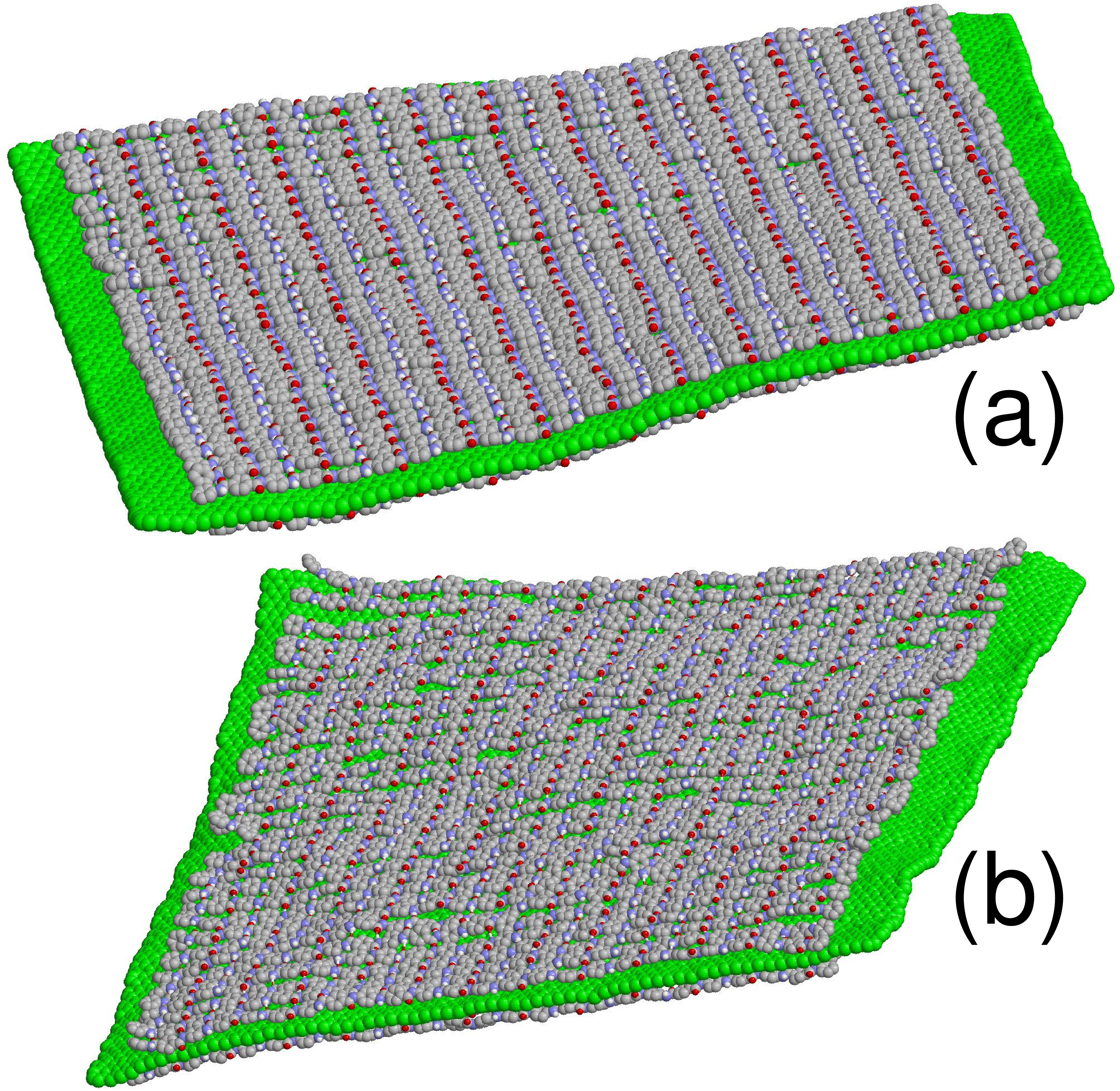}
\end{center}
\caption{\label{fig06}\protect
State of an isolated three-layer complex (Aramid$_N$)$_{N_1}$/G/(Aramid$_N$)$_{N_1}$ ($N=24$, $N_1=32$, $N_{\rm g}=12670$) at temperatures (a) $T=600$ and (b) 1200~K.
}
\end{figure}

Weakening of bonds does not lead to a change in the rectangular shape of the $\beta$-sheet -- see Fig.~\ref{fig03}.
At $T=810$~K, about $30\%$ of the bonds are weakened ($p=0.679$).
A further increase in temperature leads to the breaking of most hydrogen bonds; at $T=840$~K, most bonds are broken ($p=0.2$), and the polyglycine chain loses the shape of a flat sheet adjacent to the graphene sheet.
At $T=825$~K, a sharp jump in the dimensionless heat capacity $c(T)$ occurs, characteristic of a first-order phase transition -- see Fig.~\ref{fig02}~(b).
Here, a phase transition of the polypeptide chain from the ``flat $\beta$-sheet'' state to the ``three-dimensional coil'' state occurs -- see Fig.~\ref{fig04}.
At this temperature, the same phase transition occurs for the three-layer complex (Gly)$_N$/G/(Gly)$_N$ that does not interact with a stationary substrate.

Thus, the performed numerical modeling of dynamics shows that flat $\beta$-sheets of polyglycine chains (Gly)$_N$ placed on a graphene sheet have high stability against thermal fluctuations.
The sheets retain their shape, stabilized by parallel chains of hydrogen bonds, up to $T=800$~K.
Note that the amino acid residue Gly in protein molecules usually leads to destabilization of $\alpha$-helices and $\beta$-sheets due to its high flexibility and the absence of side chain constraints.
When a polyglycine chain is placed on a flat graphene sheet, these properties of glycine, on the contrary, lead to increased stability of the $\beta$-sheet.
Here, the high flexibility of such a chain allows it to easily fold into $\beta$-sheets, and the absence of large side chain amino acid residues allows the chain to closely adhere to the surface of the graphene sheet.
All this leads to a ``dimensionality reduction'' effect of the polyglycine chain placed on a graphene sheet.

\section{Layer of Kevlar molecules}

On a graphene sheet, structures with parallel hydrogen bond chains (\ref{f1}) can also form parallel linear Kevlar molecules (para-aramid -- polyterephthalamide) H(C$_6$H$_4$CONH)$_N$C$_6$H$_5$ -- see Fig.~\ref{fig05}.
Analysis of the Kevlar molecular structure shows that the high elastic modulus of Kevlar fibers stems from the rigidity of the aromatic polyamide chains and the high density of interchain hydrogen bonds \cite{Wang2020}.
According to \cite{Chowdhury2018}, hydrogen bonds account for approximately $60\%$ of the transverse strength of Kevlar fiber.
We will show that placing Kevlar molecules on a graphene sheet leads to a significant increase in the stability of hydrogen bonds.

We take the same graphene sheet G=C$_{12670}$ of size $17.56\times 18.58$~nm$^2$.
We place on the sheet a system of 32 parallel para-aramid molecules H(C$_6$H$_4$CONH)$_{24}$C$_6$H$_5$.
Such a two-layer structure can be described by the formula (Aramid$_N$)$_{N_1}$/G$|$, where the number of links in one para-aramid molecule (number of PGs) is $N=24$, the number of molecules is $N_1=32$, and the number of carbon atoms in the graphene sheet is $N_{\rm g}=12670$.
Such a structure is stabilized by $N$ parallel chains of hydrogen bonds of PGs (\ref{f1}).
Neighboring chains have opposite directions, and the total number of hydrogen bonds is $N_{\rm hb}=N(N_1-1)=744$.
The presence of $N+1$ flat benzene rings in each chain leads to a strong interaction of the molecule (due to $\pi$-stacking) with the flat graphene sheet \cite{Lian2014}.
A structure two times smaller with $N=14$, $N_1=16$, $N_{\rm g}=3518$ ($N_{\rm hb}=210$) is shown in Fig.~\ref{fig05}.
In addition to the two-layer structure lying on a flat substrate, we also consider an isolated three-layer structure (Aramid$_N$)$_{N_1}$/G/(Aramid$_N$)$_{N_1}$, where $N_1$ para-aramid molecules are located on each side of the graphene sheet.

To describe the dynamics of para-aramid molecules, we use the same AMBER force field that we used to model the polyglycine peptide chain.
We consider CH atomic groups included in the benzene rings as united atoms with mass $M=13m_p$.
Numerical solution of the problem of minimizing the interaction energy of the two-layer molecular system (Aramid$_N$)$_{N_1}$/G$|$ (\ref{f6}) showed that in the ground stationary state, all para-aramid molecules lie in a single plane parallel to the graphene sheet at a distance of $h\approx 3.30$~\AA~ from its surface.
The molecules lie parallel to each other, forming $N$ transverse lines of hydrogen bonds -- see Fig.~\ref{fig05}.
The total number of hydrogen bonds is $N_{\rm hb}=N(N_1-1)=744$, and the average hydrogen bond energy is $E_{\rm hb}=0.179$~eV.
The reduced hydrogen bond energy in comparison with the polyglycine $\beta$-sheet arises because the benzene rings of adjacent molecules sterically hinder the close approach of the hydrogen-bonding peptide groups.

To model the dynamics of two- and three-layer structures (Aramid$_N$)$_{N_1}$/G$|$ and (Aramid$_N$)$_{N_1}$/G/(Aramid$_N$)$_{N_1}$, we first find the ground stationary state of the molecular system by numerically solving the problem of minimizing its potential energy (\ref{f6}).
Then we place the resulting stationary state of the system into a Langevin thermostat and find its thermalized state.
To do this, we numerically integrate the system of equations of motion (\ref{f7}) with the initial condition corresponding to the stationary state of the molecular structure.
After reaching a state in equilibrium with the thermostat, we find the average energy of the system $\bar{E}(T)$ and the average number of hydrogen bonds $\bar{N}_{\rm hb}(T)$ at temperature $T$ (we assume that two peptide groups form a hydrogen bond if their interaction energy $E>0.1$~eV).
Next, using formula (\ref{f8}), we find the dimensionless heat capacity of the system $c(T)$, where the total number of atoms in the molecular system is $N_a=N_1(10N+6)$, and the fraction of PGs participating in the formation of hydrogen bonds
$$
p(T)=\bar{N}_{\rm hb}(T)/N_{\rm pg},
$$
where the total number of PGs is $N_{\rm pg}=NN_1$.

Numerical integration of the system of equations of motion (\ref{f7}) showed that two- and three-layer structures (Aramid$_N$)$_{N_1}$/G$|$ and (Aramid$_N$)$_{N_1}$/G/(Aramid$_N$)$_{N_1}$ are stable against thermal fluctuations at all considered temperatures $T\le 1600$~K -- see Fig. \ref{fig06}.
The dimensionless heat capacity of the two- and three-layer structure always remains equal to unity ($c=1$).
With increasing temperature, only a slow decrease in the fraction of peptide groups participating in the formation of hydrogen bonds occurs.
Here, in contrast to the $\beta$-sheet of the polyglycine chain, destruction of the planar structure of parallel hydrogen bond chains of PGs does not occur.
This is due to the stronger interaction of Kevlar molecules with the graphene sheet and, consequently, a stronger manifestation of the ``dimensionality reduction'' effect of the molecular system when placed on a flat graphene sheet.

The obtained results allow us to conclude that the addition of graphene to Kevlar fibers can significantly increase their thermal stability and resistance to transverse loads.

\section{Conclusion}

The performed numerical simulation of the dynamics of planar two- and three-layer molecular structures has shown that $\beta$-sheets of polyglycine peptide chains (Gly)$_N$ placed on a graphene sheet on one or both sides have high stability against thermal fluctuations.
The layers retain their shape with parallel chains of hydrogen bonds up to a temperature of $T=800$~K.
The high flexibility of the polyglycine chain allows it to easily fold into a $\beta$-sheet, and the absence of large side chain amino acid residues allows it to closely adhere to the surface of the graphene sheet.
All this together leads to a ``dimensionality reduction'' effect for the chain placed on a graphene sheet.

This effect is even stronger for a system of parallel Kevlar (para-aramid) molecules placed on the surface of a graphene sheet, due to the strong $\pi$-stacking interaction of the aromatic rings of the molecules with the sheet.
Here, parallel chains of hydrogen bonds between the peptide groups of neighboring molecules are preserved even at higher temperatures.
The performed modeling of such two- and three-layer structures allows us to conclude that the addition of graphene to Kevlar fibers can significantly increase their thermal stability and resistance to transverse loads.
\\ \\
{\bf Acknowledgements}\\

Computational facilities were provided by the Joint Supercomputer center (JSCC) of the National Research Center ``Kurchatov Institute''.
The research was funded by the Russian Science Foundation (RSF) (project No. 25-73-20038).

\end{document}